\documentclass[manuscript]{aastex}

\usepackage{graphicx,amsmath,booktabs,threeparttable,url}

\usepackage[breaklinks]{hyperref}
\usepackage[hyphenbreaks]{breakurl}

\usepackage{CJKutf8}
\usepackage{threeparttable}
\usepackage[usenames,dvips]{color}
\usepackage[normalem]{ulem}

\newcommand{\cntext}[1]{\begin{CJK*}{UTF8}{bsmi}#1\end{CJK*}}

\usepackage{natbib}

\begin{document}

\title{EVOLUTION OF POST-IMPACT REMNANT HELIUM STARS IN TYPE~Ia SUPERNOVA REMNANTS WITHIN THE SINGLE-DEGENERATE SCENARIO }

\author{Kuo-Chuan Pan (\cntext{潘國全})$^{1}$, Paul M. Ricker$^1$, and Ronald E. Taam$^{2,3}$}
\affil{$^1$Department of Astronomy, University of Illinois at Urbana$-$Champaign, 1002 West Green Street, Urbana, IL 61801, USA; kpan2@illinois.edu, pmricker@illinois.edu}
\affil{$^2$Department of Physics and Astronomy, Northwestern University, 2145 Sheridan Road, Evanston, IL 60208, USA; r-taam@northwestern.edu}
\affil{$^3$Academia Sinica Institute of Astronomy and Astrophysics, P.O. Box 23-141, Taipei 10617, Taiwan}


\begin{abstract}

The progenitor systems of Type~Ia supernovae (SNe~Ia) are still under debate. 
Based on recent hydrodynamics simulations, non-degenerate companions in the single-degenerate scenario (SDS) should survive the supernova impact.
One way to distinguish between the SDS and the double-degenerate scenario is to search for the post-impact remnant stars (PIRSs) in SN~Ia remnants.     
Using a technique that combines multi-dimensional hydrodynamics simulations with one-dimensional stellar evolution simulations,
we have examined the post-impact evolution of helium-rich binary companions in the SDS. 
It is found that these helium-rich PIRSs (He~PIRSs) dramatically expand and evolve to a luminous 
phase ($L\sim 10^4 L_\odot$) about 10~years after a supernova explosion. 
Subsequently, they contract and evolve to become hot blue-subdwarf-like (sdO-like) stars by releasing 
gravitational energy, persisting as sdO-like stars for several million years before evolving to the helium red-giant phase.   
We therefore predict that a luminous OB-like star should be detectable within $\sim 30$~years after 
the SN explosion.  Thereafter, it will shrink and become an sdO-like star in the central 
regions of SN~Ia remnants within  star-forming regions for 
SN~Ia progenitors evolved via the helium-star channel in the SDS. 
These He~PIRSs are predicted to be rapidly rotating ($v_{\rm rot} \gtrsim 50$~km~s$^{-1}$) and 
to have high spatial velocities ($v_{\rm linear} \gtrsim 500$~km~s$^{-1}$).
Furthermore, if supernova remnants have diffused away and are not recognizable at a later stage,
He~PIRSs could be an additional source of single sdO stars and/or hypervelocity stars.   

\end{abstract}



\section{INTRODUCTION}

The catastrophic explosions of Type Ia supernovae (SNe~Ia)
are of great importance in probing the history of the universe.
It is generally believed that SNe~Ia are caused by 
thermonuclear explosions of carbon-oxygen (CO) white dwarfs (WDs) in close binary systems, 
but their intrinsic variety and nature of their progenitor systems is still under debate 
\citep{2000tias.conf...33L, 2000ARA&A..38..191H,2012Natur.481..149R, 2012NewAR..56..122W}.

Current mainstream progenitor scenarios include the single-degenerate scenario
\citep[SDS;][]{1973ApJ...186.1007W, 1982ApJ...257..780N} and the double-degenerate scenario
\cite[DDS;][]{1984ApJS...54..335I, 1984ApJ...277..355W}.
In the SDS, the binary companion is a non-degenerate companion which could include stars of 
many different stellar types, including main-sequence (MS) stars, red giants (RGs), 
helium (He) stars, and M-dwarfs 
\citep{1999ApJ...522..487H, 2004ApJ...601.1058I, 2008ApJ...679.1390H, 2009MNRAS.395..847W, 2010MNRAS.401.2729W, 2010MNRAS.404L..84W, 2012ApJ...758..123W}.   
However, these non-degenerate companions are usually hydrogen- or helium-rich.
The observational upper limit of stripped hydrogen after SN impact is $< 0.01M_\odot$ 
for SN~2005am and SN~2005cf \citep{2007ApJ...670.1275L}, 
and $<0.001M_\odot$ for SN~2011fe \citep{2013ApJ...762L...5S}.
Thus, the absence of the hydrogen that should appear in SN~Ia spectra poses a problem for the SDS.
The DDS instead results from mergers of two CO WDs with total mass greater than the Chandrasekhar mass,
avoiding the hydrogen problem.  
These violent events may lead to accretion-induced collapse to neutron stars instead of thermonuclear explosions \citep{1985ApJ...297..531N}. 
However, estimates of the delay time distribution (DTD) based on observed supernova rates are consistent with a large fraction of events being due to double degenerate progenitors \citep{2011MNRAS.412.1508M, 2012MNRAS.426.3282M}.

Based on recent hydrodynamics simulations of SN~Ia impact on binary companions in the SDS, 
including grid-based \citep{2000ApJS..128..615M, 2010ApJ...715...78P, 2012ApJ...750..151P}
and smooth particle \citep[SPH;][]{2008A&A...489..943P, 2012A&A...548A...2L} simulations, 
these non-degenerate companions should survive the SN impact and could be detectable. 
Thus, one of the simplest ways to distinguish between the SDS and DDS is to search for post-impact remnant stars (PIRSs) in Type Ia supernova remnants (Ia~SNRs).  

Recent PIRS searches have studied two Galactic Ia~SNRs 
(SN~1572, \citealt{2004Natur.431.1069R, 2007PASJ...59..811I, 2009ApJ...701.1665K, 2009ApJ...691....1G, 2012arXiv1210.2713K}, 
and SN~1006, \citealt{2012Natur.489..533G, 2012ApJ...759....7K}) 
and two Ia~SNRs in the Large Magellanic Cloud \citep[LMC;][]{2012ApJ...747L..19E, 2012Natur.481..164S}. 
So far, only M-dwarf stars and the subgiant Tycho G star have emerged as possible PIRS candidates, 
and they are not well matched with companion models in the standard SDS channels, 
a result that may favor the DDS. 
However, the properties of a PIRS could change significantly after the SN impact. 
For instance, \cite{2000ApJS..128..615M} and \cite{2012ApJ...750..151P} have shown that almost all the envelope of  the RG in the RG-WD channel should be removed during the SN impact,
and $\sim 10-20\%$ of the MS star mass should be stripped and ablated in the MS-WD channel \citep{2000ApJS..128..615M, 2012ApJ...750..151P, 2012A&A...548A...2L}. 
Therefore, the PIRS in the RG-WD channel could be a helium pre-WD or low mass helium burning star
with little hydrogen-rich envelope. 
We note that the SN impact will not only strip and ablate the mass of a non-degenerate companion star, 
but also compress and deposit energy into it \citep{2012ApJ...750..151P}. 

The evolution of PIRSs has been studied by \cite{2003astro.ph..3660P} for a $1 M_\odot$ 
sub-giant companion and by \cite{2013ApJ...765..150S} for a $1 M_\odot$ MS companion. 
They found that PIRSs could be over-luminous due to the energy release from the SN energy deposition, 
suggesting that the SDS should be ruled out for several Ia~SNRs. 
However, in their calculations, these authors assumed  
ad-hoc prescriptions for energy input and mass stripping
without performing detailed hydrodynamical calculations and, therefore, did not accurately calculate the 
shock compression in the stellar interior and the depth of the energy deposition. 
In \cite{2012ApJ...760...21P}, we studied the evolution of PIRSs 
using a detailed treatment of SN impact via three-dimensional hydrodynamics simulations.
These simulations included the symmetry-breaking effects of orbital motion, rotation of the 
non-degenerate companions, and Roche-lobe overflow for the MS-WD channel. 
These three-dimensional simulation results were mapped 
into a one-dimensional stellar evolution code to simulate the post-impact evolution. 
It was found that MS-like PIRSs evolve to become subgiants ($L \sim 10-100 L_\odot$) after a few hundred years and could be slowly rotating after stellar expansion.
Although the model closest to the Tycho G star in these calculations was twice as bright, 
these results provide some support for Tycho G as a possible PIRS in the SDS. 

A new subclass of sub-luminous SNe~Ia, namely Type Iax supernovae (SNe~Iax), 
recently has been proposed by \cite{2012arXiv1212.2209F}. 
This population could originate from the He-WD channel in the SDS via a helium double-detonation explosion  or by merger of a He WD with a CO WD \citep{2012arXiv1212.2209F, 2013arXiv1301.1047W}.    
The He-WD channel naturally explains the absence of hydrogen lines, and two SNe~Iax have shown helium lines in their spectra, suggesting that helium must be present in the progenitor systems. 
\cite{2010ApJ...715...78P, 2012ApJ...750..151P} have shown that only about $\lesssim 5\%$ of the 
mass of the helium star is lost into the SNR in the He-WD channel, an amount that is much lower than the hydrogen mass lost in the MS-WD channel and the RG-WD channel. 
The He-WD channel mainly contributes to the prompt part of the delay-time distribution in population 
synthesis studies ($\sim 45-220$~Myr, \citealt{2010A&A...515A..88W}), 
and this is consistent with the distribution of SNe~Iax, 
since no SNe~Iax have been observed in elliptical galaxies. 
Finally, several He-WD binary systems with the properties required to be SN Ia progenitors
have been observed, for example 
KPD~1930+2752 \citep{2000MNRAS.317L..41M, 2007A&A...464..299G},
CD-30 11223 \citep{2012ApJ...759L..25V, 2012arXiv1209.4740G}, and
RX~J0648.0-4418 \citep{2009Sci...325.1222M}. 
These systems have either a massive WD ($M_{\rm WD} \gtrsim 1.3M_\odot$) or a short orbital period ($P_{\rm orb} < 0.05$~d). 
The helium nova V445~Pup is also considered to be a likely progenitor system \citep{2008ApJ...684.1366K}. 

In this paper, we follow the methods in \cite{2012ApJ...760...21P} and extend the study of post-impact evolution to helium-rich PIRSs (He~PIRSs) in the SDS. 
We show that He~PIRSs become hot blue-subdwarf-like (sdO-like) stars after the SN impact 
and should be detectable in Ia~SNRs.  
Hot blue subdwarfs are core helium-burning stars, and about half of them are in binary systems.
The formation of single hot blue sub-dwarfs is still an open question \citep{2009ARA&A..47..211H}, 
and the He-WD channel could be one source. He~PIRSs may also 
contribute to the hypervelocity star (HVS) population \citep{2010Ap&SS.329..293W}, 
since the original orbital speed is very high at the time of the SN explosion in the He-WD channel.
HVSs have extremely high space motions and can be unbound in our Galaxy;
for example, US~708 has $v \sin i = 708 \pm 15$~km~s$^{-1}$ \citep{2005A&A...444L..61H}. 
We consider four different helium star progenitor models from the He-WD channel in  \cite{2009MNRAS.395..847W} (HeWDa, HeWDb, HeWDc, and HeWDd in \cite{2010ApJ...715...78P}, hereafter Paper~I) and perform three-dimensional hydrodynamics simulations of the impact of SN Ia ejecta using the method described in \cite{2012ApJ...750..151P} (hereafter Paper~II).
Based on the hydrodynamics results, we carry out post-impact 
stellar evolution simulations using a method similar to the one 
described in \cite{2012ApJ...760...21P} (hereafter Paper~III).
In the next section, the numerical codes and methods are described. 
Section~\ref{sec3} gives a detailed description of the He star progenitor systems in our calculations.   
In Section~\ref{sec4}, we present the simulation results and the possible evolutionary 
tracks of He~PIRSs. We discuss several effects on the post-impact evolution and the 
possibility of He~PIRSs as sources of single hot blue subdwarfs and/or HVSs in 
Section~\ref{sec5}. In the final section, we summarize our simulation results and conclude. 


\section{NUMERICAL CODES AND METHODS \label{sec2}}

The simulation codes used in this paper are essentially the same as those in Papers~II~\&~III, 
including the three-dimensional hydrodynamics code 
FLASH\footnote{\url{http://flash.uchicago.edu}} version~3 \citep{2000ApJS..131..273F, 2008PhST..132a4046D}, 
to simulate the impact of SN Ia ejecta on the binary companions, 
and the stellar evolution code, MESA\footnote{\url{http://mesa.sourceforge.net}} 
(Modules for Experiments in Stellar Astrophysics; \citealt{2011ApJS..192....3P, 2013arXiv1301.0319P}), 
to create the progenitor models and simulate the post-impact evolution. 
We perform the SN Ia explosion simulations and post-impact stellar evolution
calculations using a technique similar to the one described in Paper~II, 
but we focus on the He-WD channel.
To link FLASH's output with MESA's initial stellar models, 
we solve the hydrostatic equilibrium equation using a fourth-order Runge-Kutta solver 
with adaptive stepsize control, as described in Paper~III. 
However, the assumption of ideal gas plus radiation equation of state (EOS) is no longer valid for He star companions, 
where the central density and temperature are much higher than in main sequence-like stars. 
We therefore update the EOS solver to include 
the OPAL, SCVH, and HELM EOS tables from the {\tt eos} and {\tt kap} modules in MESA.


\section{HE STAR PROGENITOR SYSTEMS \label{sec3}}

\begin{figure}
\epsscale{0.5}
\plotone{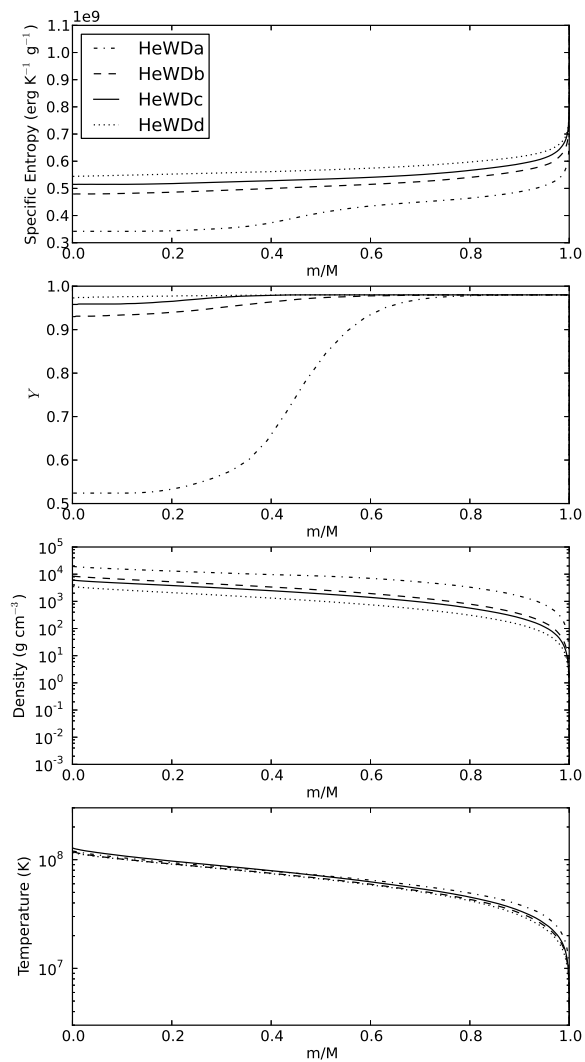}
\caption{\label{fig_models_before}
The initial specific entropy, helium composition ($Y$), density, and temperature profiles as functions
of the fractional mass before the SN~Ia 
explosion for models HeWDa, HeWDb, HeWDc, and HeWDd in Table~\ref{tab1}.}
\end{figure}

The progenitor models are taken from the HeWDa, HeWDb, HeWDc, and HeWDd helium star models in Paper~I, 
but without the simplification of uniform composition. 
These four models were generated with initial masses equal to 1.25, 1.35, 1.4, and 1.8 $M_\odot$ and initial metallicity $Z=0.02$.
An artificial constant mass loss rate was adopted such that the evolution times and final helium-star masses were consistent with the detailed binary evolution models of \cite{2009MNRAS.395..847W}.
Once the mass-loss phase ended, the stellar models were taken as initial models for the three-dimensional FLASH simulations.
The physical properties of these four initial models are summarized in Table~\ref{tab1}.
All four helium star models were relaxed on the three-dimensional grid by artificially damping the gas velocity for five dynamical timescales, ensuring that our models started in hydrostatic equilibrium and reducing the geometrical distortion introduced by passing from one dimension to three dimensions. 
The entropy, composition, density, and temperature of these four helium stars at the onset of the 
SN~Ia explosion are shown in Figure~\ref{fig_models_before}. 

\begin{table}
\begin{center}
\caption{The progenitor models at the onset of the SN explosion \label{tab1}}
\begin{tabular}{cccccc}
\\
\tableline
Model & $M_0$ ($M_\odot$) \tablenotemark{\dagger} & $R_0$ ($10^{10}$~cm) & $\log_{10} L_0$ ($L_\odot$) & $\log_{10} T_{{\rm eff}, 0}$ (K) & $P_{{\rm orb}, 0}$ (sec) \\
\tableline
HeWDa & 0.697 & 0.63 & 1.30 & 4.61 & 1,033 \\
HeWDb & 0.803 & 1.10 & 1.62 & 4.57 & 2,257\\
HeWDc & 1.007 & 1.35 & 2.11 & 4.64 & 2,682 \\
HeWDd & 1.206 & 1.61 & 1.09 & 4.35 & 3,410 \\
\end{tabular}
\tablenotetext{\dagger}{The mass ($M_0$), radius ($R_0$), luminosity ($L_0$), effective temperature ($T_{{\rm eff}, 0}$), and orbital period ($P_{{\rm orb}, 0}$) for different He star progenitor models at the time of the SN explosion, using the final masses in  \cite{2009MNRAS.395..847W}.}
\end{center}
\end{table}

\begin{figure}
\epsscale{0.8}
\plotone{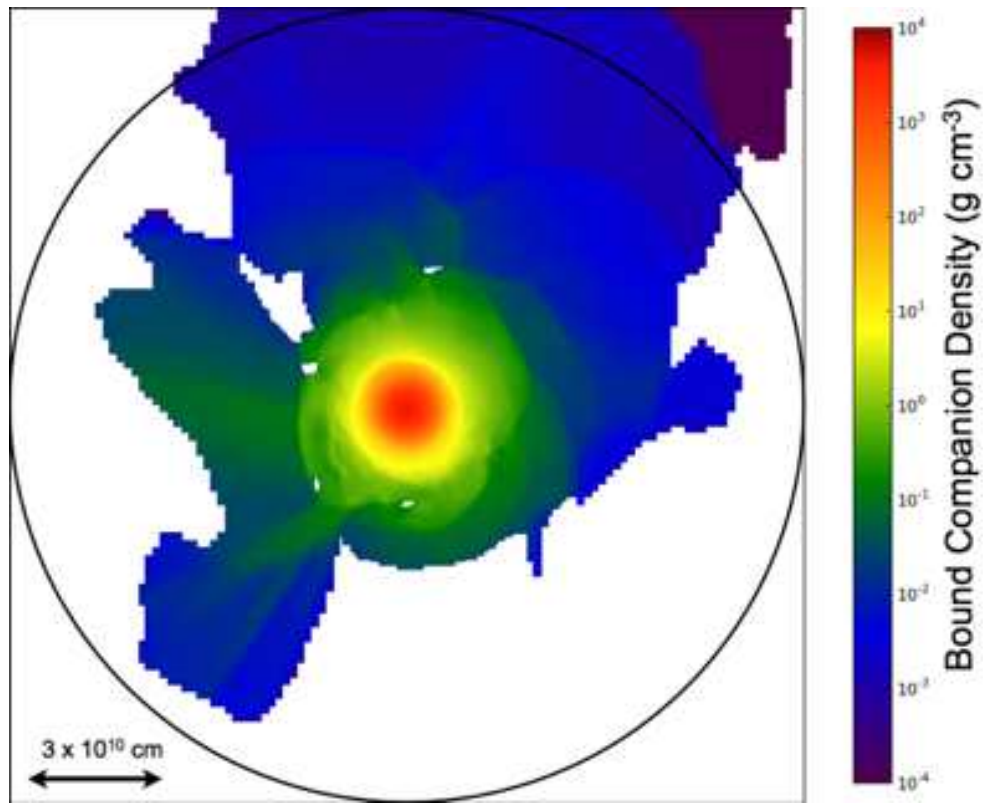}
\caption{\label{fig_flash}
Bound gas density distribution in the orbital plane for a three-dimensional SN~Ia simulation with the model 
HeWDc in Table~\ref{tab1} at time 3106 sec after the explosion. 
The black circle shows the maximum distance ($r_{\rm max} = 8.83\times 10^{10}$~cm)
which is used for the 
one-dimensional model reconstruction. }
\end{figure}
\begin{figure}
\epsscale{1.0}
\plotone{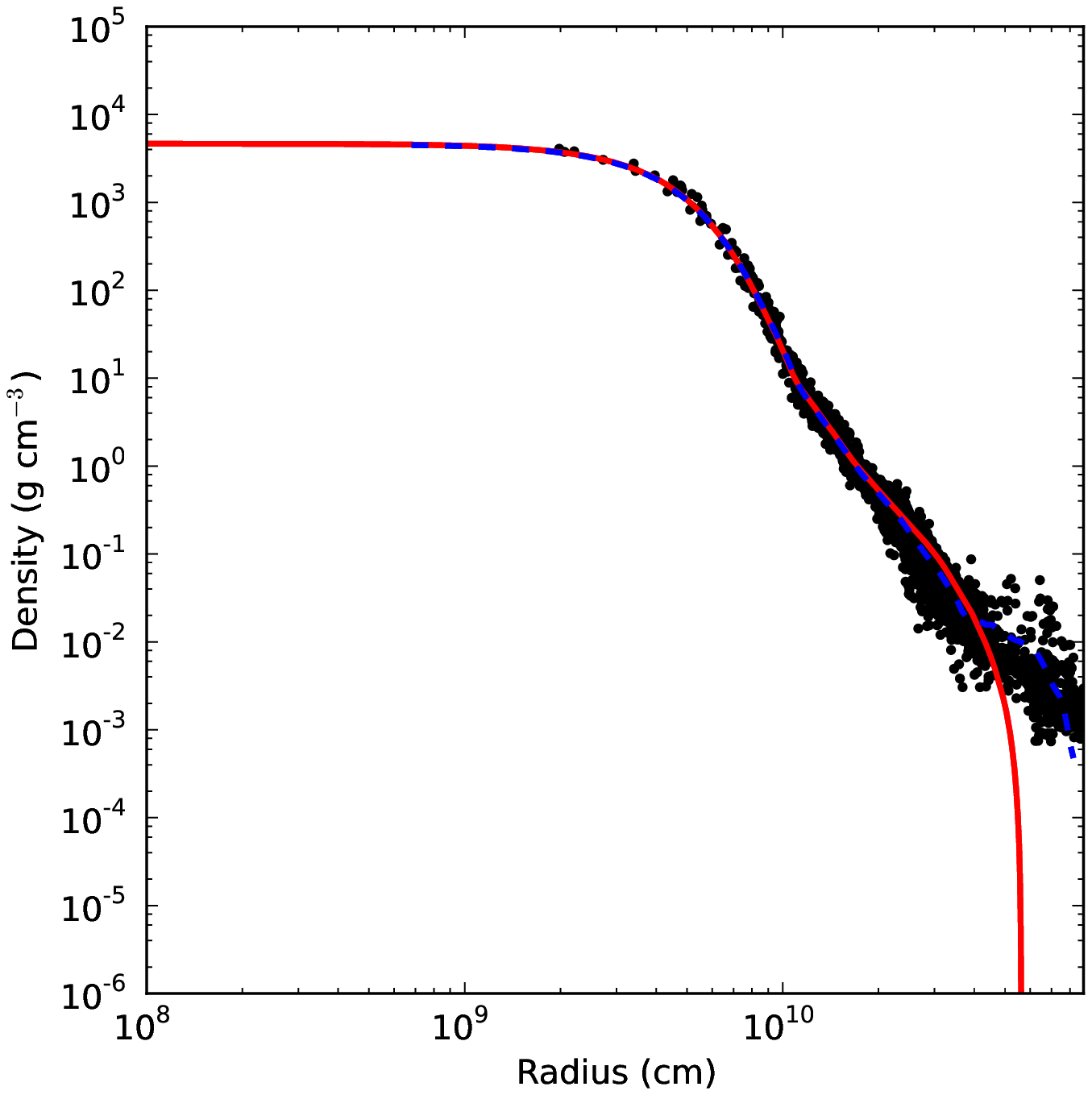}
\caption{\label{fig_radius}
Bound gas density versus radius distribution for the model HeWDc in Table~\ref{tab1} 
at time 3106 sec after the explosion.
Black dots represent the scattered values before angle-averaging (randomly selected $1/512$ of total data points).
The blue dashed line represents the angle-averaged radial profile. 
The solid red line indicates the relaxed hydrostatic solutions in MESA. }
\end{figure}

Using the numerical setup and initial conditions for SN~Ia explosions described in Paper~II, 
we performed three-dimensional FLASH simulations of SN~Ia explosions in close binary systems with resolutions of 6/8 AMR levels (equivalent to a $1024^3$ uniform grid and a zone size of $0.029 R_*$; see Paper~II for definition). 
In Paper~II, we performed a convergence study and concluded that this resolution would 
be sufficient to adequately describe the properties of post-impact companions.
The initial binary systems were assumed to be in Roche-lobe overflow (RLOF), 
and the SN model used was the W7 model \citep{1984ApJ...286..644N}.
Although SNe~Iax are mostly sub-Chandrasekhar mass explosions, we chose the standard W7 explosion as the 
first case for comparison with our previous studies in Papers~I and II.  
Since the mass loss from the non-degenerate star is sensitive to the detailed numerical setup of 
the SN model,
we adopted the ``W7 SN'' sub-grid model, which has a power-law density 
distribution and a constant temperature distribution in radius that matches the exploding mass and energy in the W7 model (see Paper~II for detailed descriptions).   

After the SN~Ia impact, the helium star binary companion loses about $\sim 4 - 6\%$ of its mass and 
is heated, the degree to which depends on the progenitor model (see Table~\ref{tab2}). 
In addition, because of the shock interaction during the SN~Ia impact, 
the central density of the helium star decreases to $\sim 20-30 \%$ of its original value, 
and the central temperature decreases by $\sim 10-15 \%$, except for model HeWDa. 
Before the SN impact, the HeWDa model has a lower central helium composition than the other three models 
due to central helium burning, causing a higher positive entropy gradient at about $m/M_* \sim 0.5$. 
Therefore, in this model the central density decreases $22\%$ during the SN impact, 
while the central temperature increases by $12\%$.

\begin{figure}
\epsscale{0.5}
\plotone{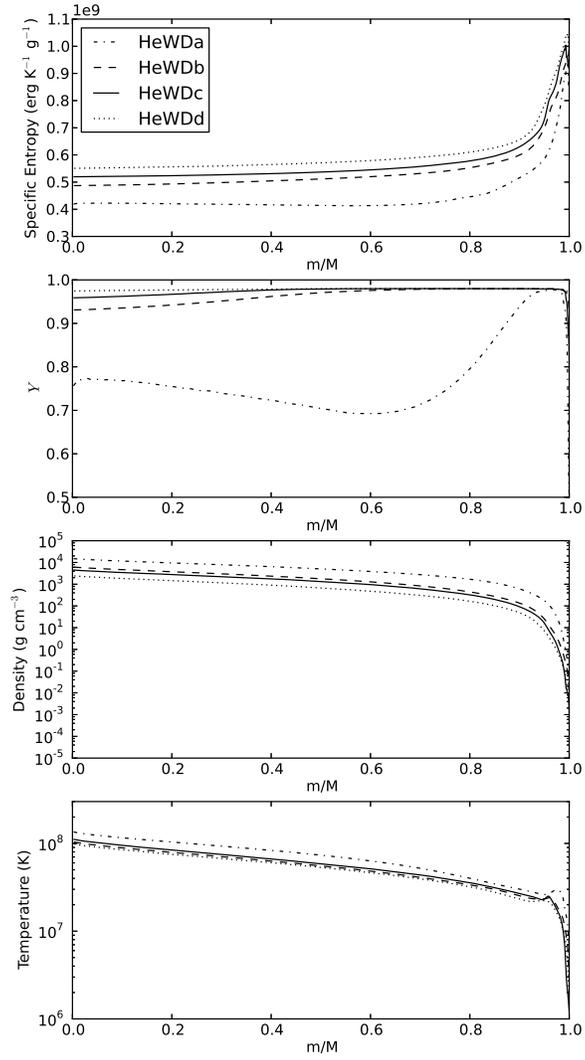}
\caption{\label{fig_models_after}
Similar to Figure~\ref{fig_models_before} but for post-impact angle-averaged profiles at the end of the FLASH simulations.}
\end{figure}

Figure~\ref{fig_flash} shows the bound gas density distribution for model HeWDc in the orbital plane at
the end of the simulation ($t= 3,106$~sec). 
Since the companion star has a high linear speed and will eventually reach the edge 
of the simulation box,
we cannot simulate the evolution for a sufficiently long time for the companion to be  
fully in hydrostatic equilibrium.  
However, we can assume the specific entropy and composition profiles are conserved if the system is adiabatic. 
To calculate the specific entropy and composition profiles, a large spherical region is considered 
(the black circle in Figure~\ref{fig_flash}) 
and then divided into $128-256$ radial bins. 
Figure~\ref{fig_radius}  
shows the comparison of angle-averaged bound density radial profile with 
the scattered 3D data points, and relaxed hydrostatic profiles for the model HeWDc. 
It is clear that the angle-averaged density profile is consistent 
with our reconstructed hydrostatic model which is described in the next paragraph. 
Figure~\ref{fig_models_after} shows the angle-averaged one-dimensional radial post-impact profiles of 
the specific entropy, helium composition, density and temperature at the end of the FLASH simulations.

With the {\tt eos} (equation of state) and {\tt kap} (opacity) module in MESA, 
we used the post-impact specific entropy profiles 
and composition profiles to construct hydrostatic models by solving the continuity and hydrostatic equations
for the density $\rho$, pressure $P$, and radius $r$ as functions of the enclosed mass $m$:
\begin{equation}
\frac{dr}{dm}=\frac{1}{4\pi r^2\rho}
\end{equation}
\begin{equation}
\frac{dP}{dm}=-\frac{Gm}{4\pi r^4}\ .
\end{equation} 
The entropy of each mass element was kept fixed, except that the entropy profiles were flattened 
in the outermost region ($0.995 < m/M_* < 1$) to avoid negative 
entropy gradients.  For the HeWDa model, the composition profile was adjusted to a uniform distribution 
due to the strong mixing during the SN impact and to avoid negative entropy and composition gradients.
The hydrostatic solutions were taken as initial conditions for the models used in MESA. 
It should be noted that the SN ejecta and unbound companion mass are optically thick at the end 
of the FLASH simulations, and therefore, the photosphere is larger than the bound companion region.
However, the outflowing ejecta will become transparent in the optical waveband and the surviving 
star should be detectable after several months. This early influence of the dynamically moving photosphere 
associated with the unbound material is not considered here.

The initial luminosity profile for MESA was estimated using the radiative temperature gradient expression,
\begin{equation}
L(m)= - \frac{(4\pi r^2)^2 ac}{3 \kappa}\frac{dT^4}{dm}\ ,
\end{equation}
where $\kappa$ is the opacity, $a$ is the radiation constant, and $c$ is the speed of light.
Since our initial luminosity profile is based on an assumption of radiative equilibrium 
and our surface profiles are not as sharp as the standard surface profile in MESA, 
the calculated photospheric luminosity and effective temperature were very rough 
at the beginning and needed to be relaxed in MESA.
A fixed time step, $\Delta t=10^{-8}$ years, was enforced for the first $10^{-6}$ years to relax the models.
After $10^{-6}$ years, we allowed the time step to be automatically determined in MESA. 

\begin{figure}
\epsscale{0.5}
\plotone{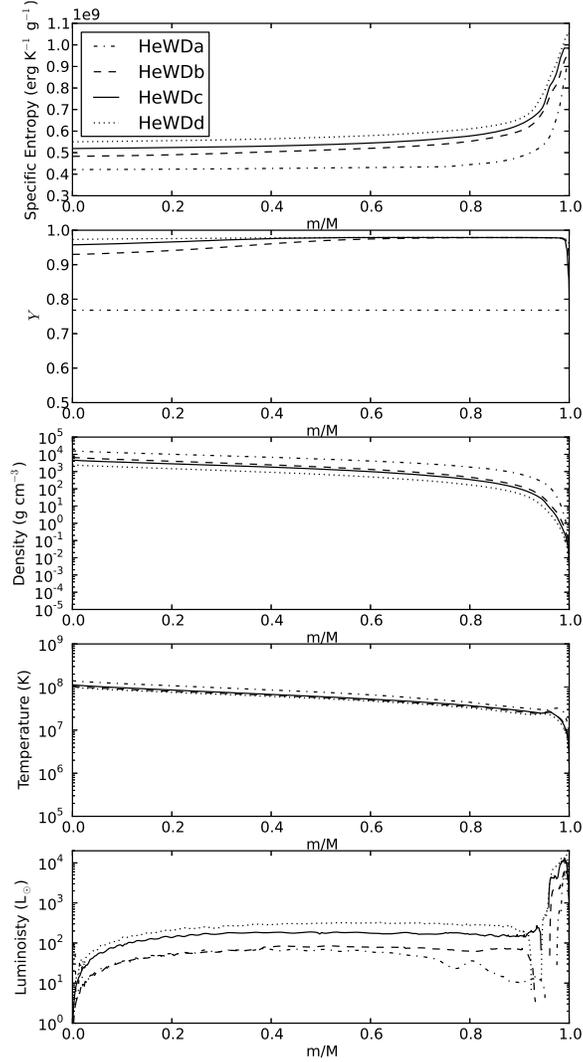}
\caption{\label{fig_models_mesa}
Relaxed post-impact companion models in MESA for specific entropy, helium composition ($Y$), density, temperature, and enclosed luminosity as functions of the fractional mass 
for models HeWDa, HeWDb, HeWDc, and HeWDd in Table~\ref{tab2}.}
\end{figure}

\begin{figure}
\epsscale{1.}
\plotone{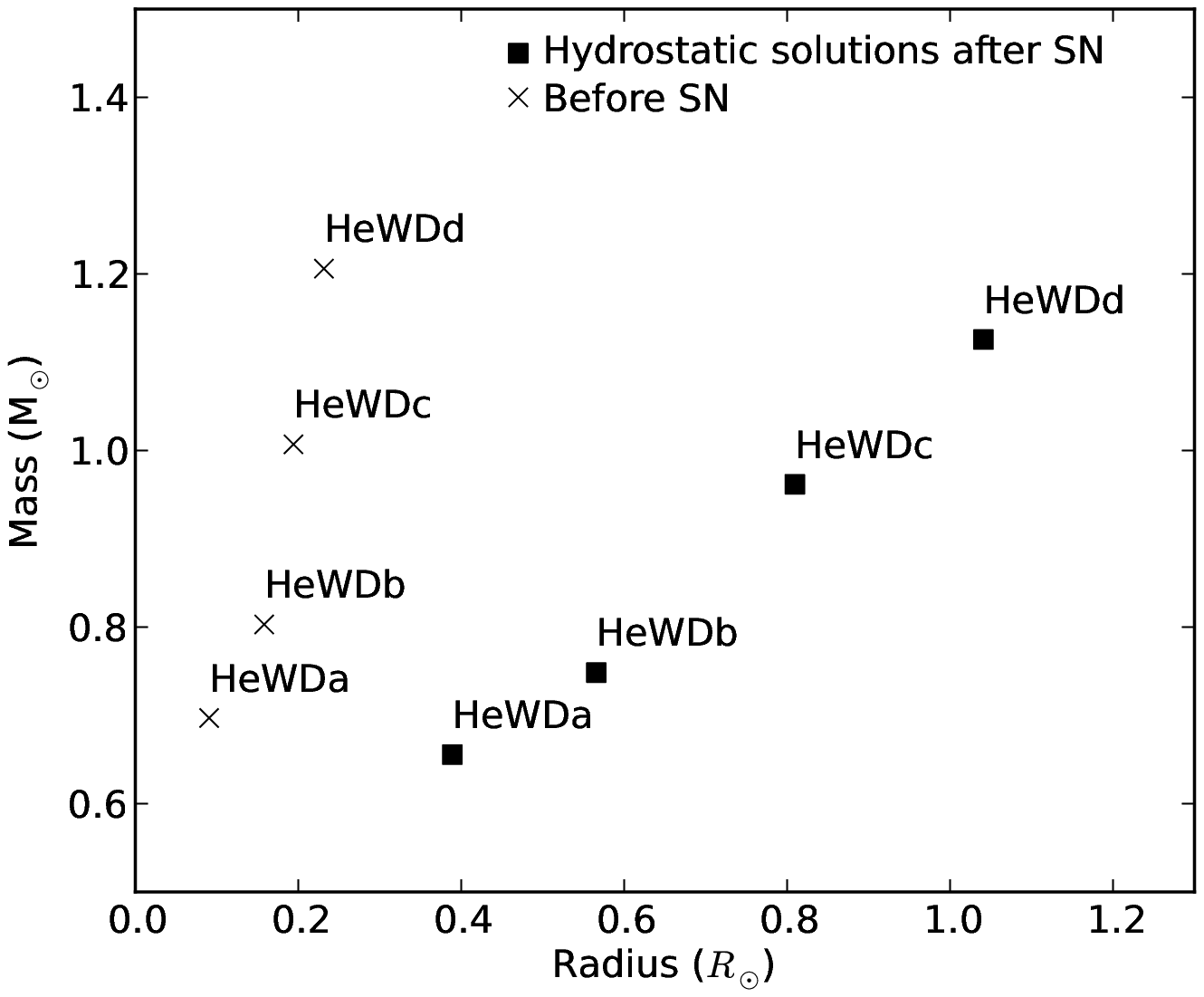}
\caption{\label{fig_mr}
Stellar mass and radius for all considered helium star models before (crosses) and after (filled squares) supernova impact. }
\end{figure}

Figure~\ref{fig_models_mesa} shows the specific entropy, helium composition, density, 
and temperature profiles of the relaxed helium star models in MESA.
The relaxed hydrostatic helium star models differ somewhat in the photospheric luminosity and 
effective temperature, but the stellar radius and interior density and temperature profiles are 
nearly the same as in the original models.  
Figure~\ref{fig_mr} shows the changes of mass and radius before and after SN impact.
Although He~PIRSs only lose a few percent of their masses, 
post-impact radii increase by a factor of $\sim 4$ (see Table~\ref{tab2}). 
These changes in radius dramatically alter the He~PIRSs. 

\begin{table}
\begin{center}
\setlength{\tabcolsep}{0.02in} 
\caption{The progenitor models immediately after the SN impact \label{tab2}}
\begin{tabular}{cccccccccc}
\\
\tableline
Model & $M_{\rm SN}$ \tablenotemark{\dagger} & ($\Delta M/M_0$)& $R_{\rm SN}$ & ($R_{\rm SN}/R_0$)& $\log_{10} L_{\rm SN}$ & $\log_{10} T_{{\rm eff, SN}}$ & $v_{{\rm linear, SN}}$ & $M_{\rm Ni}$ & $E_{\rm in}$ \\
  &  ($M_\odot$) &  & ($10^{10}$~cm) &  & ($L_\odot$) & (K) & (km~sec$^{-1}$) & $(10^{-4} M_\odot)$ & ($10^{49}$ erg) \\
\tableline
HeWDa & 0.656 & 5.88 \% & 2.71 & 4.30 & 0.265 & 4.03 & 734 & 15.0  & 1.3\\
HeWDb & 0.748 & 6.85 \% & 3.93 & 3.58 & 0.510 & 4.01& 550 & 2.38 & 1.3 \\
HeWDc & 0.962 & 4.47 \% & 5.63 & 4.17 & 0.792 & 4.01 &  509 & 5.64 & 1.5 \\
HeWDd & 1.126 & 6.6 3\% & 7.24 & 4.50 & 1.16 & 4.03 &  446 & 1.75 & 1.7\\

\end{tabular}
\tablenotetext{\dagger}{The mass ($M_{\rm SN}$), mass change ($\Delta M \equiv M_{\rm SN}-M_0$ ), radius ($R_{\rm SN}$), luminosity ($L_{\rm SN}$), effective temperature ($T_{{\rm eff}, {\rm SN}}$), linear 
spatial velocity ($v_{{\rm linear}, {\rm SN}}$), mass of bound nickel ($M_{\rm Ni}$), and energy deposition from the SN ejecta ($E_{\rm in}$) of initial relaxed post-impact hydrostatic models in MESA.}
\end{center}
\end{table}

\section{POST-IMPACT EVOLUTION \label{sec4}}

Once we have the relaxed hydrostatic stellar models, 
the post-impact evolution of He~PIRSs can be easily calculated in MESA.
In this section, we describe this evolution. 
Hertzsprung-Russell (H-R) diagrams of post-impact evolutionary tracks of He PIRSs 
are presented in luminosity versus effective temperature, surface gravity versus effective 
temperature, and color-magnitude forms, providing diagnostics for searches in future 
observations for He~PIRSs in Ia~SNRs.

\subsection{Evolutionary Tracks}

Similarly to the MS-like PIRSs in Paper~III, the He PIRSs expand rapidly and dramatically 
due to the release of energy deposited by the SN impact. 
The amount of energy deposition can be calculated by tracing the amount of binding energy increased 
after the SN impact. Since $\sim 5\%$ of the mass is lost after the SN impact, 
the initial binding energy should exclude the energy from the regions that will eventually become unbound.
We mark tracer particles that remain bound at the end of FLASH calculations and calculate
the original locations of these particles at beginning of our calculations.
It is found that the energy deposition is $1.3-1.7 \times 10^{49}$~ergs in the 
He-WD channel (see Table~\ref{tab2}). 
At RLOF, the total incident energy from supernova ejecta is $3 \times 10^{49}$~ergs, 
suggesting that $40\% - 60\%$ of the energy is absorbed by the remnant helium stars. 
The remaining incident energy takes the form of 
the kinetic and thermal energy of the unbound mass of the companion, and the 
kinetic energy due to supernova kick.  Furthermore, the reverse shock during the 
supernova impact also carries away some of the energy (see Paper~II).
 
\begin{figure}
\epsscale{0.5}
\plotone{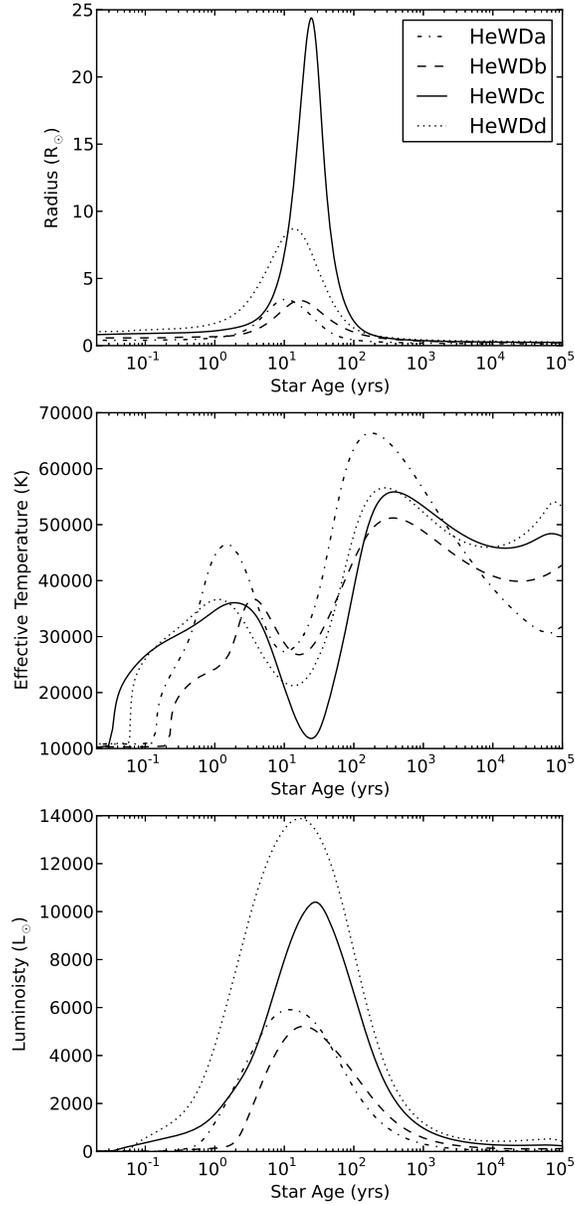}
\caption{\label{fig_evo}
Evolution of the photospheric radius, effective temperature, and luminosity as functions of time. 
Each line shows the evolution of a He~PIRS in Table~\ref{tab2}.}
\end{figure}

Figure~\ref{fig_evo} shows the evolution of the photospheric 
radius, luminosity, and effective temperature as functions of time. 
All He~PIRSs expand on a timescale of $\sim 10 - 30$ years, depending on their progenitor models. 
These expansion rates are determined by the local radiative diffusion timescale \citep{1969ApJ...156..549H}, which is associated with not only the stellar structure, but also the amount and depth of SN energy 
deposition.  Since He~stars are more compact than MS-like stars, the depths of energy deposition are 
shallower, causing a shorter local radiative diffusion timescale.  
Therefore, heat transfer initially occurs more rapidly than the thermal expansion. 
The effective temperature starts to increase at $\sim 10^{-2} - 10^{-1}$ years 
and then continues increasing for another $10^0 - 10^1$ yr up to $\sim 30,000 - 50,000$~K.
Subsequently, the surfaces of the He~PIRSs cool off to $\sim 10,000-30,000$~K on 
a timescale of $\sim 10$~years due to the expansion.
The stars quickly become luminous helium OB stars ($L \sim 10^3-10^4 L_\odot$). 
When the deposited energy has radiated away, the photospheric radii reach 
maximum values at around $\sim 10-30$~yrs. Subsequently, the stars contract and release 
gravitational energy, turning them into sdO-like stars for $\sim 10^7$~years. 
After $\sim 10^7$~years, the core helium of He~PIRSs should be exhausted, and the stars should evolve to 
the helium red giant phase. The simulations are terminated at $\sim 10^5$ years, 
since the SN~Ia remnant may not be recognizable after this time. 

We have noticed that there is $\lesssim 1 \%$ artificial mass loss during angle-averaging 
of the post-impact density, 
since the helium star is close to the edge of the simulation box at the end of the FLASH simulations. 
However, the most important factors that control the post-impact evolution are the amount of energy 
deposited and the corresponding depth, 
which are mainly concentrated in the outer $5\%$ of mass (see the temperature bump in Figure~\ref{fig_models_after}). 
We have tested the effect of mass loss by changing the mass of the post-impact remnant star and conclude that the artificial mass loss will not lead to a notable difference in the post-impact evolution. 


\begin{figure}
\epsscale{1.0}
\plotone{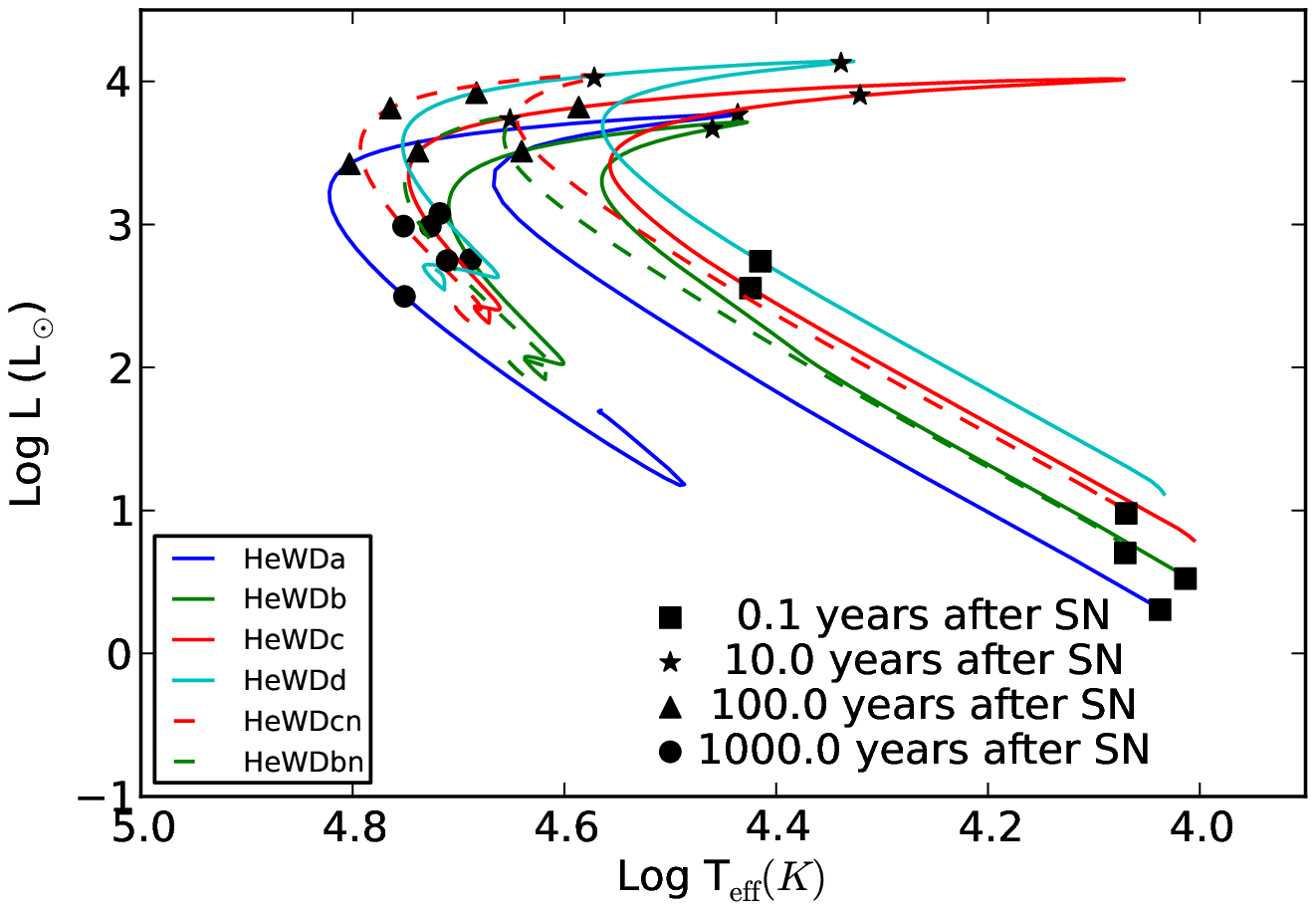}
\caption{\label{fig_tl}
Evolutionary tracks in the H-R diagram for different He~PIRS models. 
Each line represents an evolutionary track of a He~PIRS in Table~\ref{tab2} over an interval of $10^5$ years.
The filled squares indicate the conditions at 0.1 years after the SN impact; 
filled stars, 10 years after the SN impact;
filled triangles, $10^2$~years after the SN impact;
filled circles, $10^3$~years after the SN impact. 
The He~PIRS models with letter ``n'' represent the cases without nickel contamination (with dashed lines).}
\end{figure}

\begin{figure}
\epsscale{1.0}
\plotone{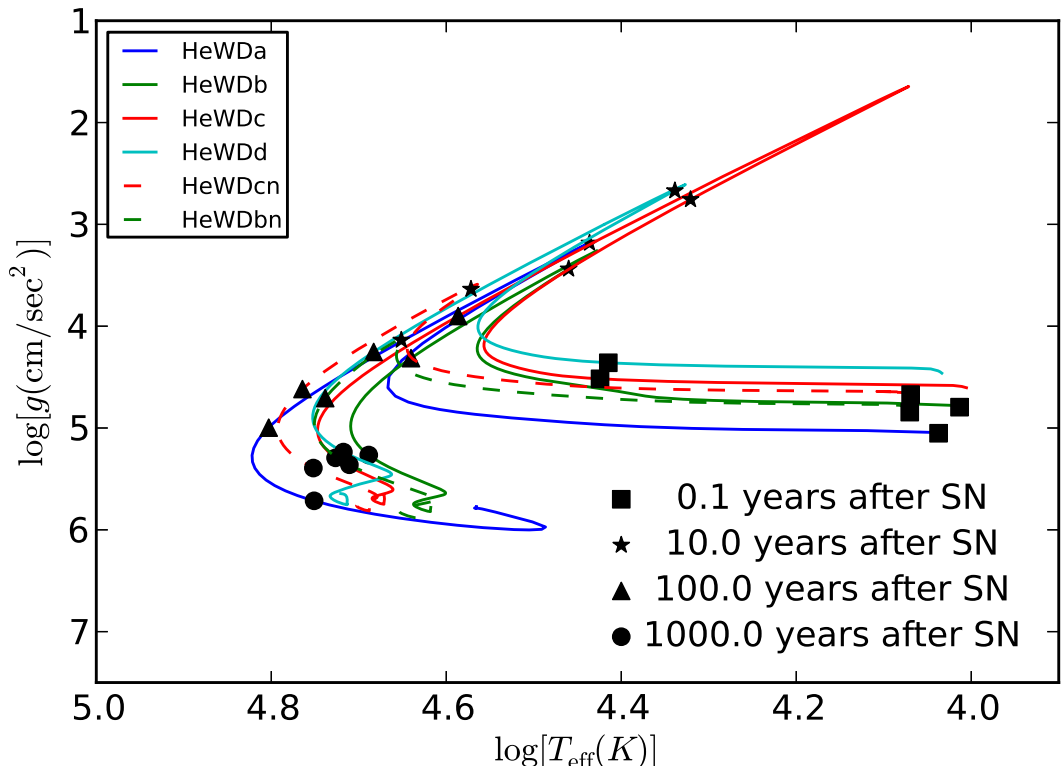}
\caption{\label{fig_tg}
Similar to Figure~\ref{fig_tl} but for surface gravity (in cgs units) versus effective temperature (K). }
\end{figure}

The evolutionary tracks of He~PIRSs in the H-R diagram are plotted in Figures~\ref{fig_tl} and \ref{fig_tg}, 
representing the effective temperature versus luminosity and effective 
temperature versus surface gravity, respectively. 
Based on our post-impact simulations, we predict that He~PIRSs will evolve to luminous OB stars on 
a timescale of $\sim 30$~years and will fade within a hundred years.
Therefore, only in young SN~Ia remnants can one observe luminous helium 
OB stars at the center of the SNR if the non-degenerate companion was a He star.
However, an sdO-like  star is observable for the remaining time. 
Note that for the He-WD channel in the SDS the delay time ($\sim 45-220$~Myr, 
\citealt{2010A&A...515A..88W}) 
is much shorter than in other SDS channels since the helium star was formed from a more massive star.  
This suggests that He~PIRSs are expected to be detected only in star-forming regions.  

As SNe~Iax represent a sub-class of sub-luminous SNe~Ia, 
their explosion energy and ejecta speed are lower than in
the standard W7 model, causing a reduced effect of the SN impact on binary companions. 
In~Paper~III, we studied the effect of the SN explosion energy on the post-impact evolution and
found that the explosion energy dramatically affects the amount and depth of SN energy depositition in
MS-like~PIRSs. For a lower explosion energy, the effect of the SN impact is shallower, 
causing a shorter radiative diffusion timescale.  
However, these differences become small once the deposited energy has radiated away. 
Therefore, for He~PIRSs with sub-luminous explosions, 
He~PIRSs behave similarly after $\sim 30$~years, but with less mass lost during the SN impact.
   


\subsection{Color-Magnitude Diagram}

\begin{figure}
\epsscale{1.0}
\plotone{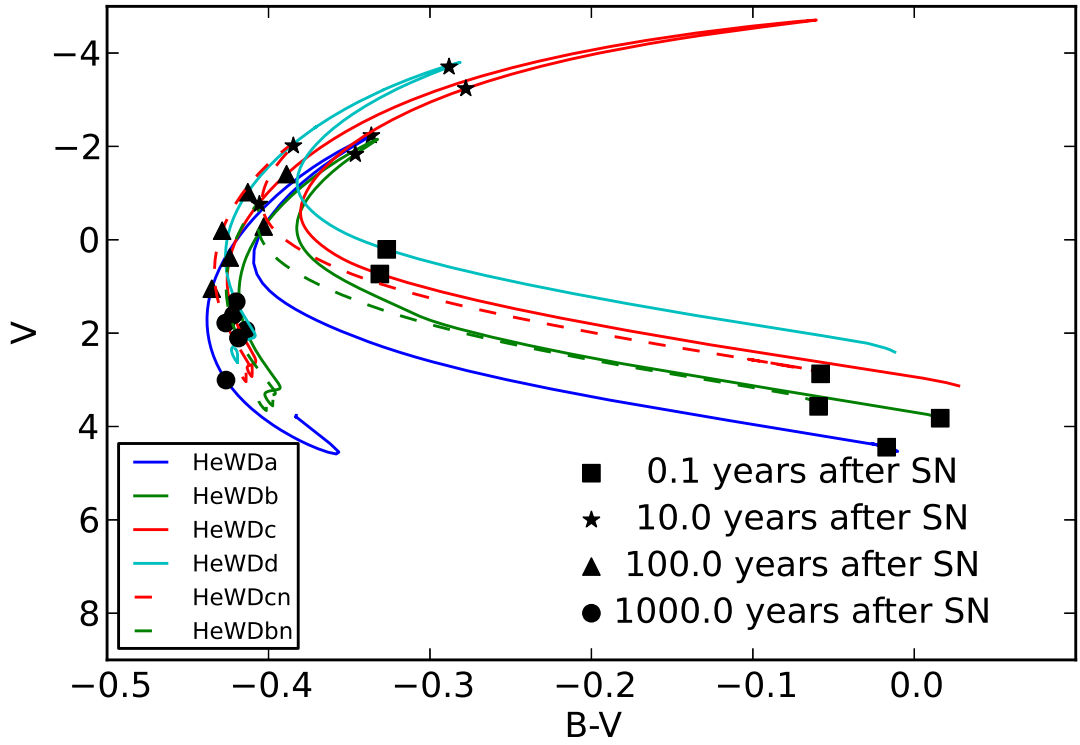}
\caption{\label{fig_cm}
Similar to Figure~\ref{fig_tl} but for V magnitude versus B-V color. The magnitudes are absolute magnitudes using the broadband u and v filters in the AB magnitude system. }
\end{figure}

For direct comparison with observations, we convert the luminosity to magnitude in the optical bands. 
Given the effective temperature ($T_{\rm eff}$) and photospheric radius ($R$) of a PIRS, the 
magnitude of the PIRS can be calculated using Equation~\ref{mag} with an assumption of blackbody radiation:  
\begin{equation}
m_{S_\lambda} = - 2.5 \log_{10} \left[ \frac{ \int S_\lambda (\pi B_\lambda) d\lambda}{ \int (f^0_\nu c/\lambda^2)S_\lambda d\lambda} \left(\frac{R}{d} \right)^2 \right] \label{mag},
\end{equation}
where $S_\lambda$ is the sensitivity function of a given filter at wavelength $\lambda$, $B_\lambda$ is the Planck function, $d$ is the distance of the PIRS, and $f^0_\nu = 3.631 \times 10^{-20}$~erg~cm$^{-2}$~s$^{-1}$~Hz$^{-1}$ is the zero-point value in the AB magnitude system.
Figure~\ref{fig_cm} shows the color-magnitude evolutionary trajectories of He~PIRSs.
The absolute magnitudes are calculated using the broadband u and v filters.  


\section{DISCUSSION \label{sec5}}

In this section, we study the possible observational effects of the post SN impact evolution.  
In particular, we examine the effect of nickel contaminations from the supernova ejecta on the 
post-impact evolution.  The surface rotational speeds of He~PIRSs during post-impact evolution are 
also predicted as functions of time. 
Furthermore, we also discuss the possibility that He~PIRSs could be a source of single sdO stars and/or HVSs.  

\subsection{Nickel Contamination}

\begin{figure}
\epsscale{0.5}
\plotone{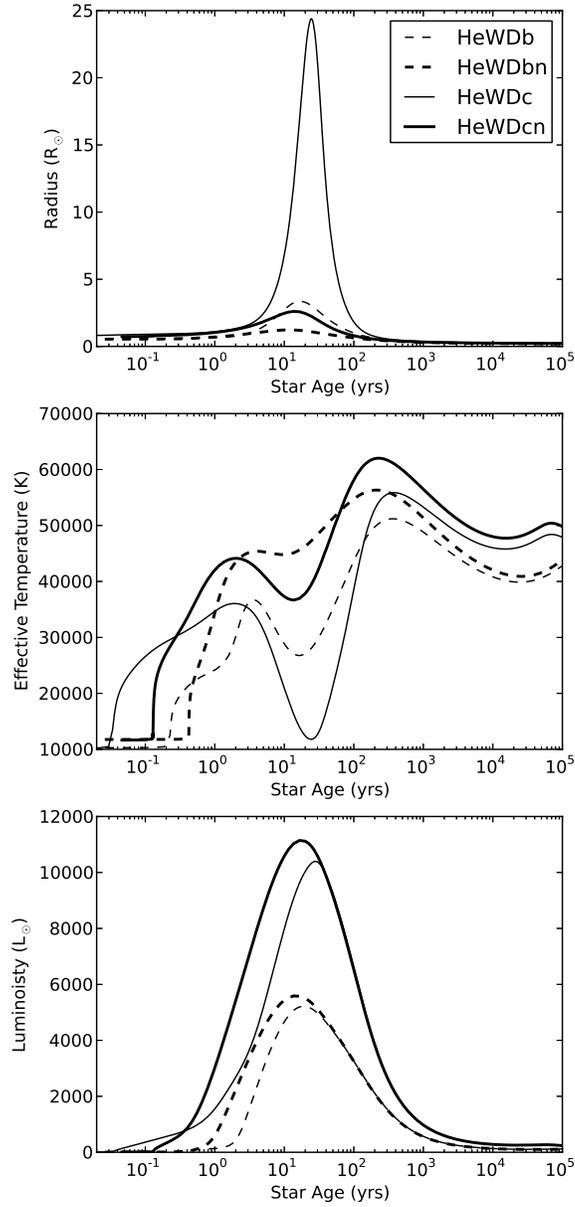}
\caption{\label{fig_nickel}
Similar to Figure~\ref{fig_evo}, but for HeWDb and HeWDc with and without nickel contaminations.
The letter ``n'' in He~PIRS models represents the cases without nickel contamination. 
}
\end{figure}

The envelope of a companion star could be contaminated by the supernova ejecta during the supernova impact or as fallback. 
In Paper~II, we have shown that this nickel contamination is much greater for He~PIRSs ($1.7 \times 10^{-4} M_\odot - 1.5 \times 10^{-3} M_\odot$; see Table~\ref{tab2}) than MS-like~PIRSs ($< 10^{-5} M_\odot$).
We note that the nickel contamination can affect the post-impact evolution for He~PIRSs, since the change of 
metallicity in the envelope will also change the opacity and radiative diffusion timescale. 
To study this effect, two He~PIRSs (labeled ``HeWDbn" and ``HeWDcn") without nickel contamination have been reconstructed by removing the bound nickel at the end of the FLASH simulations. 
The hydrostatic profiles of the reconstructed models do not change significantly, 
but they have slightly smaller radii. 
Figure~\ref{fig_nickel} shows the evolution of the photospheric 
radius, effective temperature, and luminosity of these two models with and without (cases with the letter ``n'') nickel contamination. 
Removing the bound nickel causes the opacity to be lower in the outer regions, making stars more transparent and causing smaller photospheric radii. 
It is clear that the maximum radii of He~PIRSs without nickel contamination are
much smaller than the same He~PIRSs with nickel contamination. 
Note that we used the fixed metal tables (Type~I) opacity module in MESA. 
Thus, the opacity contributed by heavy metals in our calculations is not very accurate, 
since the abundance ratios are different in the nickel-contaminated region than the abundances assumed in OPAL. 
However, the difference between models with and without nickel contamination is not very significant in the H-R diagrams (see Figure~\ref{fig_tl}).
Thus, these differences can be treated as an uncertainty in our post-impact simulations.   
 
Energy generation by $^{28}$Ni decay and $^{28}$Co decay in the SN ejecta-contaminated regions is 
neglected in our calculations. This is justified by the fact that the
energy generated by these nuclear decays is much smaller than the energy deposited from the 
supernova ejecta since the mass of bound nickel is small.


\subsection{Surface Rotational Speed \label{sec_rot}}

\begin{figure}
\epsscale{1.}
\plotone{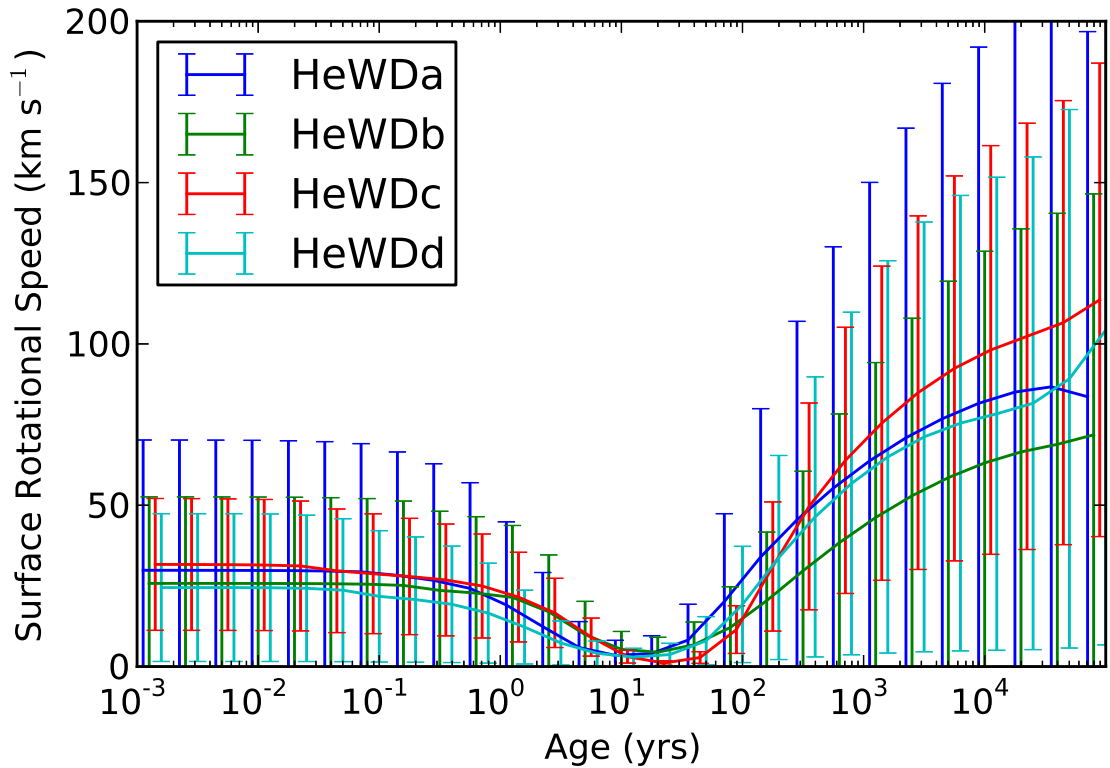}
\caption{\label{fig_vrot}
Evolution of surface rotational speed for all He-WD models in Table~\ref{tab2}. 
The error is estimated using the variation of specific angular momentum within the envelope of initial hydrostatic solutions (see Section~\ref{sec_rot} for detailed description). The number of error bars is only chosen for visualization and does not represent the actual number of data points. }
\end{figure}

In Paper~III, we showed that the surface rotational speed could decrease 
to $\sim 10-20$~km~s$^{-1}$ for MS-like~PIRSs, if the specific angular momentum is conserved during the post-impact evolution. 
This result implied that the possible PIRS candidate Tycho G could 
not be completely ruled out because of its low surface rotation speed. 
Here, we apply the same method to He~PIRSs. 
The He stars were set into rigid rotation at the beginning of the FLASH simulations because the synchronization time due to tidal locking for a given mass ratio $q$ and orbital period $P$
\citep{1977A&A....57..383Z}, $t_{\rm sync} \sim 10^4((1+q)/2q)^2P^4~{\rm yrs} \ll 1$~day, is extremely short in the He-WD channel.  
After the SN impact, a He~PIRS loses $\lesssim 10\%$ of its angular momentum and is no longer in 
a state of rigid-body rotation. 
Thus, an angle-averaged, post-impact radial angular velocity profile can be calculated by averaging different latitudes and longitudes in spherical coordinates in the FLASH output. 
The post-impact surface rotational speed can be calculated by assuming the specific angular momentum to be conserved after the SN impact.
Figure~\ref{fig_vrot} shows the post-impact evolution of the 
surface rotational speed versus evolution time.
It is found that the angle-averaged angular velocity profiles exhibit some variations in the 
envelope region, and we estimate the uncertainty by using the standard deviation of the specific angular momentum within the envelope.
Since He stars are more compact and are closer to the accreting WD at the time of the SN explosion, 
the surface rotational speed should be much higher than for MS-like companions in the MS-WD channel. 
The surface rotational speed at the time of the SN~Ia explosion could be as high as $\sim 200$~km~s$^{-1}$ if the He star companion is close to co-rotation due to tidal locking.  
For He~PIRSs, our calculations show that the surface rotational speed should decrease 
to $\lesssim 10$km~s$^{-1}$ when the star expands to its maximum size. 
However, because the radiative diffusion timescale is short, the surface rotational speed is slow only during the first hundred years and will eventually increase to $\sim 100$~km~s$^{-1}$ after a few hundred years. 
Therefore, if He~PIRSs exist in historical Ia~SNRs, their surface rotational speeds could be higher than $50$~km~s$^{-1}$, depending on the progenitor models and the ages of the SNR.


\subsection{He PIRSs as hypervelocity stars?}

During the past decade a number of HVSs have been observed in the halo of our Galaxy.  
They are sub-luminous O- or B-type stars or MS B stars with high radial velocities that could exceed the 
escape velocity of the Milky Way. 
The common explanation for their formation involves tidal ejection in binary stars (or triple stars) 
associated with either a massive black hole or binary black holes \citep{1988Natur.331..687H, 2003ApJ...599.1129Y}. 
Alternatively, \cite{2009A&A...508L..27W} suggest that the surviving He~PIRSs in SNe~Ia within the 
He-WD channel could be a source of HVSs. 
In our calculations, He~PIRSs are sdO-like stars 
for $\sim 10^7$~yr and thereafter evolve to the helium red-giant phase.
The final linear velocity (original orbital velocity plus kick velocity) of our He~PIRSs is 
$\sim 400-800$~km~s$^{-1}$ (see Table~\ref{tab2}).
He~PIRSs could move to a distance of $\sim 10$~kpc within $10^7$~years, becoming single sdO-like stars in the halo of our Galaxy, if the Ia~SNRs have diffused away at that time. 
The HVS US~708 (or HVS2) has a radial velocity $v \sin i = 708 \pm 15$~km~s$^{-1}$, 
$T_{\rm eff} = 44,500$~K, $\log g = 5.23$, and Galactic latitude $b= +47.05^\circ$ 
at a distance of 19~kpc \citep{2005A&A...444L..61H}, 
giving a displacement of 14~kpc from the Galactic plane. The effective temperature and surface gravity 
of US~708 are consistent with our He~PIRS models, but they require a subdwarf lifetime longer than 
several times $10^7$~yr to reach such a distance, if the SN exploded in the Galactic plane.
Therefore, a less massive model may better match US~708, since less massive models have higher 
linear speeds and longer subdwarf lifetimes.


\section{CONCLUSIONS}

We have investigated the post-impact evolution of He~PIRSs within the SDS for SNe~Ia via numerical simulations. 
Four helium star models from \cite{2009MNRAS.395..847W} are considered in our calculations. 
We performed three-dimensional hydrodynamics simulations using the methods that were 
described in Paper~II.
The post-impact evolution of these stars has been studied by reconstructing hydrostatic models based 
upon the final output in the hydrodynamics simulations and then interpolating them into a one-dimensional stellar evolution code, MESA.
It is found that He~PIRSs expand dramatically and evolve to become luminous OB stars within 
about $\sim 10-30$~years after the SN~Ia explosion.    
This phase is short ($< 100$~years), and therefore these luminous OB stars are not likely to be detected 
in historical Ia~SNRs. 
After $\sim 30$~years, He~PIRSs contract and evolve into hot blue-subdwarf-like (sdO-like) stars due to 
the release of gravitational energy. 
Therefore, we predict that most He~PIRSs should be sdO-like stars and could be detectable in nearby Ia~SNRs.  

We also predict that these He~PIRSs should be rapidly 
rotating ($v_{\rm rot} \gtrsim 50$~km~s$^{-1}$). 
Although a few fast rotating hot blue-subdwarfs have been observed recently \citep{2011ApJ...733L..13G, 2013arXiv1301.4129G}, 
most single hot blue-subdwarfs (sdOs and sdBs) are slowly rotating \citep{2012A&A...543A.149G}. 
If single hot blue-subdwarfs originate from the merger of two He WDs \citep[sdOs/sdBs;][]{2009ARA&A..47..211H}, 
merger via a common envelope phase on the red giant branch \citep{2008ApJ...687L..99P}, 
or the He-WD channel in SNe~Ia (sdOs), the theoretical predictions cannot explain these observations, 
suggesting that other mechanisms operate to slow down the rotation, such as magnetic braking. 

The He-WD binary channel is favored for the prompt DTD in the SDS 
and is expected to occur in star-forming regions. 
Since the orbital period immediately prior to 
the SN~Ia explosion in the He-WD channel is extremely short ($\lesssim 1$~hour), the system is 
expected to be tidally locked. 
Thus, He stars should be rapidly rotating at the time of the SN~Ia explosion. 
Although some angular momentum will be lost during the SN impact, He~PIRSs are still expected to be 
rapidly rotating after $\sim 30$~years. 
The spatial velocity of He~PIRSs is also expected to be high, reflecting the high orbital speed 
at the time of the SN~Ia explosion and implying that He~PIRSs could contribute to the HVS 
population (i.e. US~708). 


\acknowledgments

We thank the anonymous referee for his/her valuable comments and suggestions for improving this paper.
The simulations presented here were carried out using the NSF XSEDE Ranger system at the 
Texas Advanced Computing Center under allocation TG-AST040034N.  
FLASH was developed largely by the DOE-supported ASC/Alliances Center for Astrophysical Thermonuclear Flashes at the University of Chicago.  
This work was partially supported by the Computational Science and Engineering (CSE) fellowship at the University of Illinois at Urbana-Champaign.
Part of the analysis of FLASH data was completed using the analysis toolkit {\tt yt} \citep{2011ApJS..192....9T}.


\bibliography{ref}


\end{document}